\documentclass[submission]{eptcs}
\providecommand{\event}{SCAV 2018} 
\usepackage{underscore}           
\usepackage{amssymb}
\usepackage{amsmath}
\usepackage{txfonts}
\usepackage{mathdots}
\usepackage{graphicx}

\title{Identification of Risk Significant Automotive Scenarios Under Hardware Failures}
\author{Mohammad Hejase, Arda Kurt, Umit Ozguner
\institute{Department of Electrical and Computer Engineering\\ The Ohio State University\\ Columbus, Ohio, USA}
\email{hejase.1@osu.edu, ozguner.1@osu.edu}
\and
Tunc Aldemir 
\institute{Department of Mechanical and Aerospace Engineering\\ The Ohio State University\\ Columbus, Ohio, USA}
\email{aldemir.1@osu.edu}
}
\def\titlerunning{}
\def\authorrunning{M. Hejase, A. Kurt, T. Aldemir, and U. Ozguner}
\begin{document}
\maketitle

\begin{abstract}
The level of autonomous functions in vehicular control systems has been on a steady rise. This rise makes it more challenging for control system engineers to ensure a high level of safety, especially against unexpected failures such as stochastic hardware failures. A generic Backtracking Process Algorithm (BPA) based on a deductive implementation of the Markov/Cell-to-Cell Mapping technique is proposed for the identification of critical scenarios leading to the violation of safety goals. A discretized state-space representation of the system allows tracing of fault propagation throughout the system, and the quantification of probabilistic system evolution in time. A case study of a Hybrid State Control System for an autonomous vehicle prone to a brake-by-wire failure is constructed. The hazard of interest is collision with a stationary vehicle. The BPA is implemented to identify the risk significant scenarios leading to the hazard of interest.
\end{abstract}

\section{Introduction}

Emerging cars in today’s markets have tens of interconnected Electronic Control Units (ECUs) that have to realize possibly thousands of features \cite{broy2006challenges}. As the level of autonomous functions in cars keep increasing, the need for alternatives to physical testing for ensuring safe operation of these functions increases. 
Ensuring safe operation of an engineered system is accomplished by inferring conditions and causes that could lead to the violation of safety goals (safety analysis). Johansson \cite{johansson2015importance} discusses a method that ensures completeness of safety goals definition through the definition of hazardous events. The method reduces the problem of providing an assurance case that supports controller compliance to safety goals to providing proof that contributions of modeled uncertainties and behaviors only lead to hazardous events within an acceptable risk. Quantitative analysis methods are typically used for estimating likelihoods of reaching hazardous events, or violating safety goals under certain system failures. Among the most common methods for quantitative analysis in the automotive industry are quantitative Failure Mode and Effect Analysis (FMEA), quantitative Fault Tree Analysis (FTA), Markov Models, and Reliability Block Diagrams \cite{modarres2016book, verma2016probabilistic}.\\
Over the past few years, research towards the development of tools and methods that provide compliant quantitative assurance cases for autonomous vehicle features have intensified. Takeichi et al. \cite{takeichi2011failure} describe a priority FTA calculation approach for latent faults. Das and Taylor \cite{das2016quantified} demonstrate a structured and systematic quantitative FTA which shows various techniques for the calculation of fault tree metrics. Zhang et al. \cite{zhang2010model} present a case study for applying combinations of FTA and FMEA techniques for thorough model based hazard analysis of autonomous systems. Cherfi et al. \cite{cherfi2014modeling} use Markov chains to model behaviors of a large class of automotive systems protected by safety mechanisms. Hoffman and Scharfenberg \cite{hofmann2015random} show, via an example, compliance of a standard cell balancing circuit with requirements set by industry standards with respect to random failures.\\ 
Traditionally, reachability analysis has been a widely used analytical tool for verification of automated vehicle safety using simulation techniques \cite{alam2014guaranteeing, althoff2014online, lawitzky2014determining, park2014hybrid}. Reachability analysis works by computing the set of all reachable states when sensor measurements, disturbances, and initial vehicle states are uncertain. Safety is ensured by computationally confirming that none of the reachable states violate a safety goal. Authors of \cite{alam2014guaranteeing, althoff2014online, lawitzky2014determining, park2014hybrid}, utilize reachability analysis as a proof to compliance with safety goals for various scenarios and case studies. Limiting features to this type of analysis, however, as noticed in the aforementioned work, are typically challenges associated with using high fidelity nonlinear models which lead to long computation times.  It is also challenging to develop a generic approach based on reachability analysis that can be used on a wide spectrum of scenarios.\\
Hybrid system analysis techniques have been proposed for the verification of control functions in cyber-physical systems. Such methods are powerful tools as they can provide formal verification of large-scale systems. Loos et al. \cite{loos2011adaptive} developed a formal model of a distributed car control system in which an arbitrary number of vehicles sharing a highway use adaptive cruise control. The authors performed a full verification of the system by utilizing a modular proof structure. Mitsch et al. \cite{mitsch2017formal} also use hybrid state analysis to formally prove safety of robot vehicles under sensor uncertainty and actuator perturbation. Such techniques are ideal use in systems with known dynamics and behaviors. However, these techniques have challenges when incorporating random hardware failures or using high-fidelity simulators that have dynamics without explicit analytic forms, such as look-up tables.\\
Currently, software development in the automotive industry follows the V process \cite{weber2009automotive}. In the V process, testing is left to the latter stages of development. New standards are being developed that emphasize on testing in the earlier stages of design.\\ 
One common way of testing in the early design stages is done by assigning specifications to Model Based Designs (MBDs) of autonomous features, and testing using simulation techniques, including fault simulation \cite{amberkar2000system,  lu2009approach, oetjens2014safety, wilwert2005quantitative}. The use of MBDs allows for system testing via accurate simulation. This approach eliminates the high costs of testing over extensive distances in various environments and locations.\\
In this paper, a generic Backtracking Process Algorithm (BPA) algorithm is proposed for the determination of quantitative metrics that probabilistically rank scenarios leading to user specified Top Events (e.g. hazardous events) by risk significance \cite{yang2016algorithm}. Within the context of this paper, a risk significant scenario is defined as a sequence of events that lead to an undesirable consequence with probability higher than a user-specified threshold. An event is defined as a change in system dynamic behavior and configuration that occurs over time. The BPA is a deductive and memory efficient implementation of a Markov Cell-to-Cell Mapping Technique (CCMT) \cite{aldemir2010probabilistic} that is used for risk informed identification and quantification of critical scenarios leading to undesirable consequences (Top Events). Markov/CCMT allows for the global analysis of dynamic systems under both epistemic and aleatory uncertainties. Probabilistic system evolution is quantified in time, and fault propagation is traced throughout the system. Markov/CCMT has mostly been used in literature for the failure analysis and diagnostics of process control systems under uncertainties \cite{aldemiruse, aldemir1987computer,  aldemir1996process, aldemir2010probabilistic, belhadj1995cell, belhadj1991probabilistic,  dinca1999fault, yang2016algorithm}. More specifically, BPA is proposed to solve the problem of tracing fault propagation in systems with complex dynamics and varying configuration, such as random hardware failures of system components. Generally speaking, existing approaches either have challenges with accurately capturing high-fidelity system dynamics, or when incorporating possible random component failures and configuration changes. The algorithm has already been used in a validation and verification framework development for Unmanned Aircraft Systems (UAS) as part of the System-Wide Safety Assurance Technologies (SSAT) initiative taken by the National Aeronautics and Space Administration (NASA) which the authors conducted jointly with ASCA, Inc. \cite{ guarro2017formal, guarro2017risk, hejase2018dynamic, hejase2017quantitative}.\\
Sections \ref{OverviewSection} and \ref{MarkovSection} of the paper provide, respectively, overviews of the autonomous ground vehicle controller design framework the analysis in this paper is based on. Section \ref{MarkovSection} describes Markov/CCMT, and the BPA. Section \ref{CaseStudySection} presents the case study under consideration in this paper, the definition of the different possible types of brake failures, and the hazard of interest. Section \ref{SimulationSection} illustrates the use of the BPA in identifying the risk significant scenarios. Section \ref{ChallengesSection} provides a discussion on the identified challenges and future directions. Section \ref{ConclusionSection} gives the conclusions of the study.

\section{Overview of Control System Design Framework} \label{OverviewSection}

\begin{figure}[htbp]
\centering
\includegraphics[width=14.5cm]{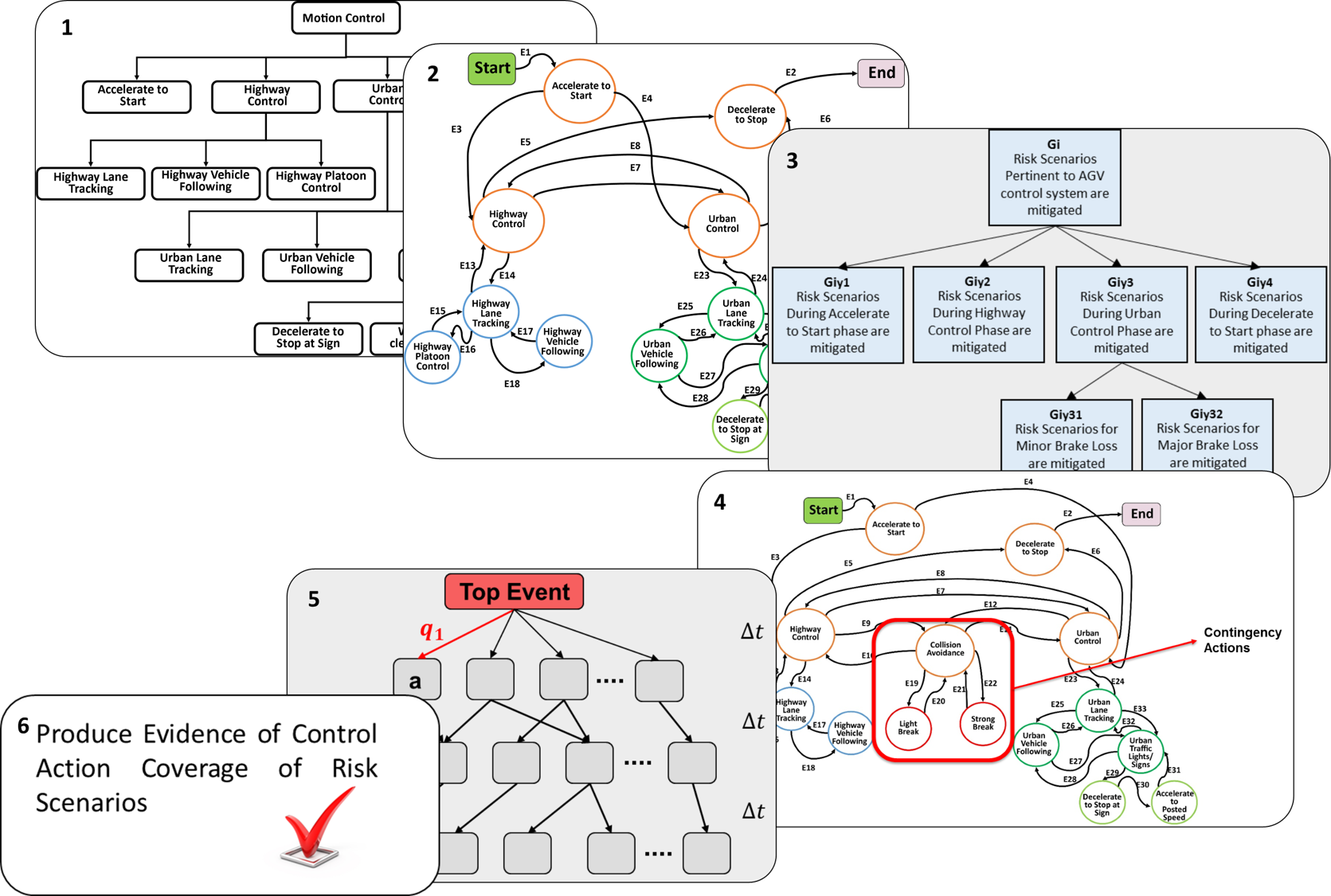}
\caption{Model Based Validation and Verification Framework (adapted from \cite{guarro2017formal, guarro2017risk})}
\label{fig1}
\end{figure}

Based on the model based validation and verification framework described in \cite{guarro2017formal, guarro2017risk} , a similar framework was constructed for the validation and verification of an autonomous ground vehicle controller. The framework, depicted in Fig. \ref{fig1}, is made up of six main elements,\\

\indent
1)	\textit{Definition of Control System Functional Hierarchy}: The control system functional hierarchy is designed based on the work of Ozguner \cite{ozguner1990coordination}. This allows for the initial definition of system and mission requirements and specifications, and the subsequent decomposition into mission phase specific functionality.\\
\indent
2)	\textit{Design of a Finite State Machine}: A Finite State Machine (FSM) representation of the high level mission controller is designed based on the control system functional hierarchy. Each state of the FSM corresponds to a different phase of the mission, and transitions between those states are determined via the definition of event based rules. \\
\indent
3)	\textit{Development of Risk Prioritized Scenarios}: The system top level safety goals are first determined. A safety case goal and evidence tree model reflecting risk prioritized scenarios encompassing nominal, contingency, and emergency conditions and actions is developed using Goal Structure Notation (GSN) \cite{kelly2004goal}. This model allows breaking up each of the top level safety goals into a set of hazards, where collectively avoiding these hazards represent the safety goals.\\
\indent
4)	\textit{FSM Augmentation with Emergency and Contingency Actions}: Based on the determined hazards, contingency and emergency actions are defined and incorporated into the FSM.\\
\indent
5)	\textit{Expansion of Risk Prioritized Scenarios for Details Analysis}: Each of the specified hazards, which can be thought of as the consequence of a risk significant scenario, is expanded upon with Markov/CCMT for detailed analysis under relevant hardware failures. The aim of this analysis is to provide an assurance case for system compliance with a target probability metric for hardware failures. Physical motion of the system is represented in this step via the use of a high-fidelity simulator. \\
\indent
6)	\textit{Produce Evidence of Control Action Coverage of Risk Scenarios}: The results produce evidence of control action coverage of prioritized operational and risk scenarios, supporting a controller assurance case.

\section{Markov Cell to Cell Mapping Technique} \label{MarkovSection}

Section \ref{MarkovOverview} presents an overview of Markov/CCMT and the required assumptions, along with the required assumptions.  In Section \ref{BPASec}, BPA is illustrated and described in detail.

\noindent 
\subsection{Overview and Assumptions} \label{MarkovOverview}

Markov/CCMT is a logic tool used to provide quantified metrics for system reliability and safety \cite{yang2016algorithm, aldemir2010probabilistic, aldemir1987computer, belhadj1995cell, belhadj1991probabilistic, aldemiruse, aldemir1996process, dinca1999fault}. Theoretical basis of BPA is presented in the work of Yang and Aldemir \cite{yang2016algorithm}. System evolution in time is represented through a series of discrete-time transitions among computational cells that partition the system state-space in a manner similar to finite element or finite difference methods. Each cell can be regarded as accounting for the uncertainty in the system location at a given point in time. A transition probability from one system cell to another is determined via system dynamics, controller behavior, or system constituent malfunction. Such transitions reflect a probabilistic mapping of the system state-space onto itself, including system hardware normal or faulted states, over a user defined time-step$\ \mathrm{\Delta}$\textit{t}.

Two assumptions are placed on the system of interest in order to employ Markov/CCMT: 

\begin{enumerate}
\item  The system components configurations are fixed over$\ \mathrm{[}t,\ t + \mathrm{\Delta}t\mathrm{)}$, but can change at$\ \mathrm{\ }t+\mathrm{\Delta}t$.

\item  Transitions among cells or hardware states do not depend on system history. 
\end{enumerate}

The first assumption means that the system components can only fail or change their mode of operation once during the interval$\ \mathrm{\Delta }t$. Through proper selection of the time-step$\ \mathrm{\Delta }t$, the system configuration changes and the probabilities of those changed can be realistically modeled and captured. The second assumption leads to the system having Markov property.  However, the second assumption can be relaxed via the use of sufficient number of auxiliary state variables.

\noindent 


\subsection{BPA} \label{BPASec}
\begin{figure}[htbp]
\centering
\includegraphics[width=14cm]{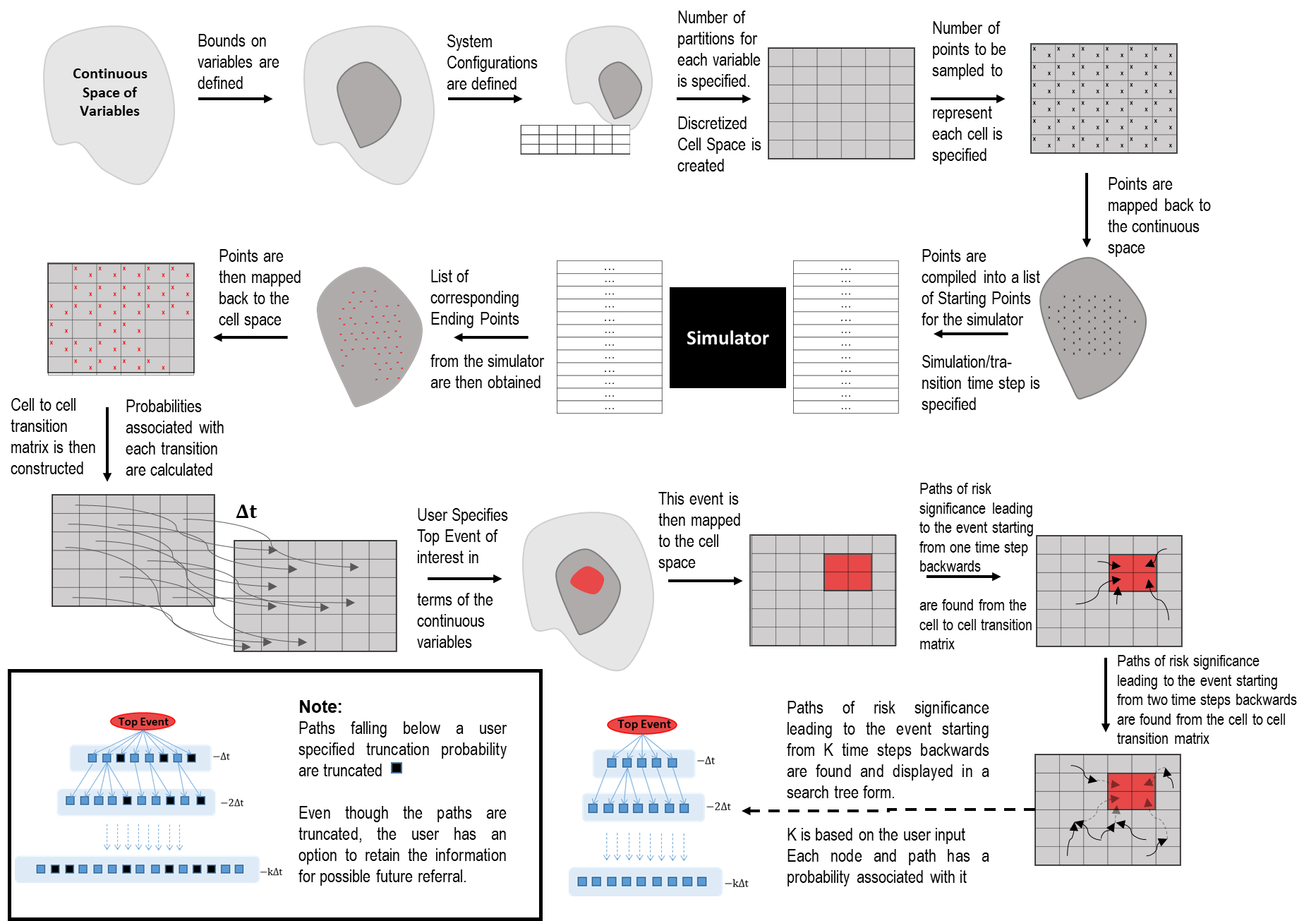}
\caption{BPA flowchart (adapted from \cite{hejase2017quantitative})}
\label{fig3}
\end{figure}

\indent
BPA is depicted in Fig. \ref{fig3}. The system continuous state-space is first discretized, and system components/configurations are defined. The combination of the discretized state-space, and the system configurations form the complete space of the system. Using a simulator, a cell-to-cell mapping of the complete space is constructed under a user-specified time-step. A Top Event of interest is specified, and sequential paths of risk significance leading to the Top Event for a user-specified search depth are identified. Probabilities are associated to each of the identified sequences.

In Section \ref{MarkovSystemDisc}, System discretization into a cell space is described. Section \ref{MarkovC2CProbCalc} contains the method of cell-to-cell transition probability calculation. An equal weight quadrature scheme is included in Section \ref{MarkovQuad}. Section \ref{RiskSigSection} describes the process for the identification of risk significant event sequences.

\subsubsection{System Discretization} \label{MarkovSystemDisc}
\indent 

The continuous$\mathrm{\ }L\mathrm{\ }$dimensional state space is represented by$\mathrm{\ }\mathcal{X}\mathrm{\triangleq }{\mathbb{R}}^L$. The$\ M$ dimensional discrete state space of the system components is represented by$\mathrm{\ }\mathcal{N}\mathrm{\triangleq }{\mathbb{Z}}^M$. The space$\mathrm{\ }\mathcal{X}\mathrm{\triangleq }{\mathbb{R}}^L$is discretized by partitioning each continuous variable $x_l\mathrm{\ (}l\mathrm{=1,\dots ,\ }L\mathrm{)}$ into intervals of$\mathrm{\ }J_l$ partitions and considering combinations of those partitions to form the cells. Knowledge of the state-space upper bounds$\ \mathrm{\ }\overline{x},$ and lower bounds$\ \mathrm{\ }\underline{x}$ is required for the partitioning. The cells can be regarded as means to accommodate epistemic uncertainties (such as model uncertainties) or aleatory uncertainties (such as process noise and minor environmental disturbances).

The possible states of each hardware component$\mathrm{\ }M$ of interest are then defined (e.g. operational, degraded, failed), with each component \textit{m}, having$\mathrm{\ }N_m$ possible states, each denoted by$\ n_m\ (m=1,\dots ,\ M)$. 

The unique combinations of the partitioned $\mathcal{X}\mathrm{\triangleq }{\mathbb{R}}^L$  along with the discrete system component configurations forms the complete state-space of the system, denoted by$\ \mathcal{V}$. Each cell in the cell space is represented by an$\mathrm{\ (}L\mathrm{+}M\mathrm{)}$ dimensional vector$\mathrm{\ }\left[\boldsymbol{\mathrm{j}}\boldsymbol{\mathrm{\ }}\boldsymbol{\mathrm{n}}\right]\mathrm{\equiv }\mathrm{[}j_1,\ \dots ,j_l,\ \cdots ,\ j_L,\ n_1,\ \dots ,\ n_m,\ \cdots ,\ n_M]$, where $\mathrm{(\ }j_l=1,2,\dots ,\ J_l;l=1,\dots ,L)$ enumerate the partitioning of the interval$\ \underline{x_l}\le x_l<\overline{x_l}\ $, and $\ n_m\ $ represents the state of component$\ \ m$ $\left(n_m=1,\dots ,N_m;\ m=1\ ,\ \dots ,\ M\right)$. The cell space$\ \mathcal{V}$ is composed of $J\times N$ unique cells with $J=\ \mathrm{\ }J_{\mathrm{1}}\mathrm{\times }\mathrm{\cdots }\mathrm{\times }J_L\mathrm{\ and\ }N=\ N_{\mathrm{1}}\mathrm{\times \dots \times }N_M$.

Let$\mathrm{\ }{\mathcal{V}}_{\mathcal{X}}\mathrm{\triangleq }{\mathbb{Z}}^L$ be a subspace of$\mathrm{\ }\mathcal{V}$ containing the vectors \textbf{j}. Let$\mathrm{\ }{\mathcal{V}}_{\mathcal{N}}\mathrm{\triangleq }{\mathbb{Z}}^M$ be a subspace of$\mathrm{\ }\mathcal{V}$ containing the vectors \textbf{n}. Note that$\mathrm{\ }{\mathcal{V}}_{\mathcal{X}}\mathrm{\cup }{\mathcal{V}}_{\mathcal{N}}\mathrm{=}\mathcal{V}$. 

\begin{figure}[htbp]
\centering
\includegraphics[width=9.5cm]{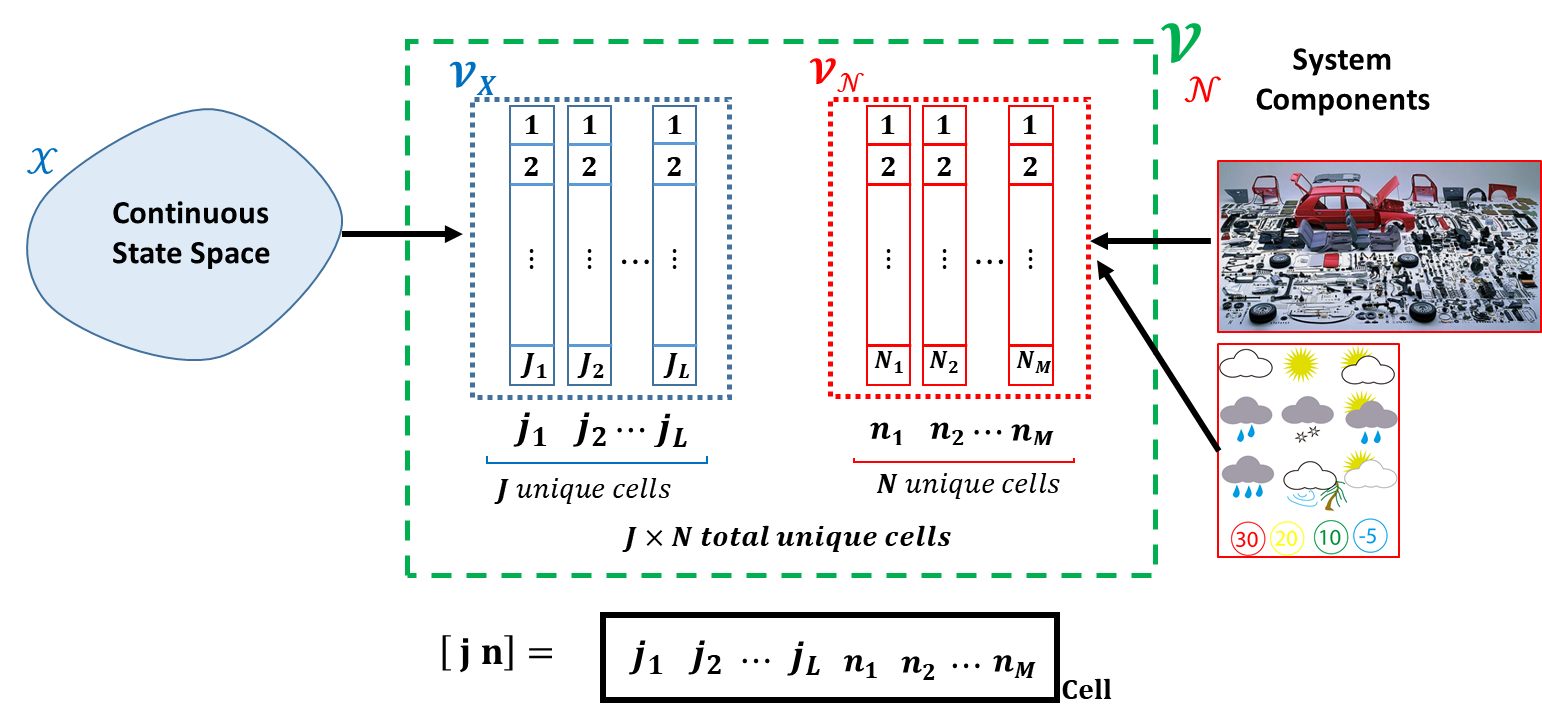}
\caption{Illustration of Discretized System}
\label{fig2}
\end{figure}

The discretized system, along with the relevant notations is illustrated in Fig. \ref{fig2}. 

\noindent

\subsubsection{Cell to Cell Transition Probability Calculation} \label{MarkovC2CProbCalc}

Using the Markov property, and as derived in [20], the cell-to-cell probabilities over a single time-step transition$\mathrm{\ }\mathrm{\Delta}$\textit{t} can be calculated from
\begin{equation}  \label{eq:3}
q\left(\boldsymbol{j},\boldsymbol{\mathrm{n}}\boldsymbol{\mathrm{|\ }}\boldsymbol{\mathrm{j}}\boldsymbol{\mathrm{',}}\boldsymbol{\mathrm{n}}\boldsymbol{\mathrm{'}}\mathrm{,\ }\mathrm{\Delta }\mathrm{t}\right)\mathrm{=}
h\left(\boldsymbol{\mathrm{n}}\boldsymbol{\mathrm{|}}\boldsymbol{\mathrm{n}}\boldsymbol{\mathrm{',\ }}\boldsymbol{\mathrm{j}}\boldsymbol{\mathrm{'}}\boldsymbol{\mathrm{\to }}\boldsymbol{\mathrm{j}},\mathrm{\ }\mathrm{\Delta}t\right)\mathrm{\times }g\left(\boldsymbol{\mathrm{j}}\boldsymbol{\mathrm{|}}{\boldsymbol{\mathrm{j}}}{\boldsymbol{\mathrm{'}}},\boldsymbol{\mathrm{n}}\boldsymbol{\mathrm{'}}\mathrm{,\ }\mathrm{\Delta}t\right)\ \ \ \  
\end{equation} 
where $g\left(\boldsymbol{\mathrm{j}}\boldsymbol{\mathrm{|}}{\boldsymbol{\mathrm{j}}}{\boldsymbol{\mathrm{'}}},\boldsymbol{\mathrm{n}}\boldsymbol{\mathrm{'}}\mathrm{,\ }\mathrm{\Delta}t\right)$represents the transition probability from cell $\mathrm{\ }\boldsymbol{\mathrm{j}}\boldsymbol{\mathrm{'}}\ $to $\boldsymbol{\mathrm{j}}\ $over$\ \mathrm{\Delta}$\textit{t} under configuration$\ \boldsymbol{\mathrm{n}}\boldsymbol{\mathrm{'}}$, and$\mathrm{\ }h\left(\boldsymbol{\mathrm{n}}\boldsymbol{\mathrm{|}}\boldsymbol{\mathrm{n}}\boldsymbol{\mathrm{',\ }}\boldsymbol{\mathrm{j}}\boldsymbol{\mathrm{'}}\boldsymbol{\mathrm{\to }}\boldsymbol{\mathrm{j}},\mathrm{\ }\mathrm{\Delta}t\right)$ quantifies the system configuration transition probabilities over$\mathrm{\ }\mathrm{\Delta }$t. 

For each component of interest $m,$ a component state transition probability matrix$\mathrm{\ }H_{n_m}$ is constructed. Contents of this matrix represent the probability of component state transitions over \( \Delta\)t. These probabilities can be based on hardware component data, such as failure rates, or expert opinion in the absence of reliable data. An example of such a matrix can be seen in Table \ref{tab:SampleMatrix} where ${\lambda }_{n'_m, n_m}$ denotes the transition rate from \( n_m' \) to \( n_m \). 

\begin{table}[htbp]
\caption{Sample System Configuration Transition Matrix ${H_{n_m}}$}
\label{tab:SampleMatrix}
\centering
\vspace{0.1cm} 
\hspace{6cm} 
Final System Configuration State\\
\begin{tabular}{llllll}
\cline{3-6}
& & Normal State & Failure State 1 & ... & FailureState N \\
& & (N)& (${F_{1}}$) & ... & (${F_N}$)  \\ \cline{1-6}
Initial System & Normal State (N) & ${\lambda_{N,N}} \Delta t $ & ${\lambda_{N,F_1} \Delta t}$ & ... & ${\lambda_{N,F_N} \Delta t}$ \\
Configuration State & Fail 1 State (${F_{1}}$) & ${\lambda_{F_1,N}} \Delta t $ & ${\lambda_{F_1,F_1} \Delta t}$ & ... & ${\lambda_{F_1,F_N} \Delta t}$ \\
 & ${\vdots}$ & ${\vdots}$ & ${\vdots}$ & - & ${\vdots}$ \\
 & Fail N State (${F_{N}}$) & ${\lambda_{F_N,N}} \Delta t $ & ${\lambda_{F_N,F_1} \Delta t}$ & ... & ${\lambda_{F_N,F_N} \Delta t}$ \\
\end{tabular}
\end{table}

  Using the Chapman-Kolmogorov equation under the assumptions stated earlier, the system cell-to-cell state transition probabilities$\mathrm{\ }g\left(\boldsymbol{\mathrm{j}}\boldsymbol{\mathrm{|}}{\boldsymbol{\mathrm{j}}}{\boldsymbol{\mathrm{'}}},\boldsymbol{\mathrm{n}}\boldsymbol{\mathrm{'}}\mathrm{,\ }\mathrm{\Delta}t\right)\mathrm{\ }$over a single time-step can be found from \cite{aldemir1987computer}
\begin{equation} \label{eq:4} 
g\left(\boldsymbol{\mathrm{j}}\boldsymbol{\mathrm{|}}{\boldsymbol{\mathrm{j}}}{\boldsymbol{\mathrm{'}}},\boldsymbol{\mathrm{n}}\boldsymbol{\mathrm{'}}\mathrm{,\ }\mathrm{\Delta}t\right)\mathrm{=}\frac{\mathrm{1}}{v_{\boldsymbol{j}\boldsymbol{'}}}\int_{v_{\boldsymbol{j}\boldsymbol{'}}}u_{\boldsymbol{\mathrm{j}}}\left[\boldsymbol{\mathrm{x}}\boldsymbol{\mathrm{(}}{\boldsymbol{\mathrm{x}}}{\boldsymbol{\mathrm{'}}},{\boldsymbol{\mathrm{n}}}{\boldsymbol{\mathrm{'}}},\mathrm{\Delta}t\boldsymbol{\mathrm{)}}\right]{dx'}
\end{equation} 
\begin{equation} \label{eq:5} 
u_{\boldsymbol{\mathrm{j}}}\left[\boldsymbol{\mathrm{x}}\boldsymbol{\mathrm{(}}{\boldsymbol{\mathrm{x}}}{\boldsymbol{\mathrm{'}}},{\boldsymbol{\mathrm{n}}}{\boldsymbol{\mathrm{'}}},\mathrm{\Delta}t\boldsymbol{\mathrm{)}}\right]~\mathrm{=}~\left\{ \begin{array}{c}
 \begin{array}{c}
\mathrm{1\ }~~~~if~\mathrm{x}\mathrm{\in }v_{\boldsymbol{\mathrm{j}}}~ \\ 
\mathrm{0\ \ }otherwise \end{array}
 \end{array}
\right. 
\end{equation} 
where $v_{\boldsymbol{\mathrm{j}}}$ is the volume of the cell$\ \boldsymbol{\mathrm{j}}$,
\begin{equation} \label{eq:6} 
\boldsymbol{\mathrm{x}}\boldsymbol{\mathrm{(}}{\boldsymbol{\mathrm{x}}}{\boldsymbol{\mathrm{'}}},{\boldsymbol{\mathrm{n}}}{\boldsymbol{\mathrm{'}}},\mathrm{\Delta}t\boldsymbol{\mathrm{)}}\mathrm{=}~\int^{t\mathrm{+}\mathrm{\Delta }t}_t{f\left(\boldsymbol{\mathrm{x}}\left(t'\right),~{\boldsymbol{\mathrm{n}}}{\boldsymbol{\mathrm{'}}}\right)}\boldsymbol{d}{\boldsymbol{t}}{\boldsymbol{'}}\mathrm{+}\boldsymbol{\mathrm{x}}\boldsymbol{\mathrm{'}} 
\end{equation} 
and$\ f\left(\boldsymbol{\mathrm{x}}\left(t'\right),~{\boldsymbol{\mathrm{n}}}{\boldsymbol{\mathrm{'}}}\right)\ $represents the equations describing system dynamics.

\subsubsection{Equal Weight Quadrature Approximation Scheme} \label{MarkovQuad}

When it is not practical or possible to evaluate \eqref{eq:6}, an equal weight quadrature approximation scheme can employed via the use of a high fidelity simulator. System location in the state space is assumed to be uniformly distributed within each cell. Multiple points are sampled to represent each cell, and are passed to the simulator to compute transitions over$\ \mathrm{\Delta}$\textit{t}. Then \eqref{eq:4} can be approximated as
\begin{equation} \label{eq:7} 
g\left(\boldsymbol{\mathrm{j}}\boldsymbol{\mathrm{|}}{\boldsymbol{\mathrm{j}}}{\boldsymbol{\mathrm{'}}},\boldsymbol{\mathrm{n}}\boldsymbol{\mathrm{'}}\mathrm{,\ }\mathrm{\Delta}t\right)\mathrm{=}     \frac{\#~of~sampled~points~in~\mathrm{cell\ }{\boldsymbol{\mathrm{j}}}{\mathrm{'}}~arriving~in\ cell\ \boldsymbol{\mathrm{J}}~over~\mathrm{\Delta }t~}{\#~of~points~sampled~from~\mathrm{\ cell\ }{\boldsymbol{\mathrm{j}}}{\mathrm{'}}\mathrm{\ }} 
\end{equation}

\subsubsection{Identification of Risk Significant Event Sequences} \label{RiskSigSection}
\indent 

While, in principle, backtracking can be accomplished through
\textbf{
\begin{equation} \label{eq:8} 
{\boldsymbol{\mathrm{\textbf{P}}}}^{\boldsymbol{\mathrm{k}}}\boldsymbol{\mathrm{=}}{\left[{\boldsymbol{\mathrm{\textbf{Q}}}}^{\boldsymbol{\mathrm{T}}}\boldsymbol{\mathrm{\textbf{Q}}}\right]}^{\boldsymbol{\mathrm{-}}\boldsymbol{\mathrm{1}}}{\boldsymbol{\mathrm{\textbf{Q}}}}^{\boldsymbol{\mathrm{T}}}{\boldsymbol{\mathrm{\textbf{P}}}}^{\boldsymbol{\mathrm{k}}\boldsymbol{\mathrm{+}}\boldsymbol{\mathrm{1}}} 
\end{equation} 
}
The BPA avoids the challenges associated with \eqref{eq:8} by using the search tree that is obtained from a probabilistic map of the system state-space onto itself. This search tree structure is achieved by recursive enumeration of sub-trees emanating from Top Event in decreasing time and the traversal of possible paths through a branching process. In order to avoid a numerical catastrophe, only risk significant scenarios with probabilities above a user-specified cut-off value are identified. 


In this study, an undesirable event$\ \mathcal{E}$ is assumed to be defined through the specification of event upper bounds$\mathrm{\ }\overline{e}$, event lower bounds$\mathrm{\ }\underline{e}$ in terms of the continuous variables in the state-space, and the set of event system configurations$\mathrm{\ }e_s$ as stated in \eqref{eq:9}-\eqref{eq:11}: 
\begin{equation} \label{eq:9} 
\overline{e}\mathrm{=}\left\{\left[\overline{e_{\mathrm{1}}}\mathrm{,\ }\mathrm{\cdots }\mathrm{,\ }{\overline{e}}_l,\ \cdots \overline{e_L}\right]\mathrm{\ }\right|\mathrm{\ }\overline{e_l\mathrm{\ }}\mathrm{\le }\overline{x_l}\mathrm{,\ }\overline{e_l}\mathrm{>}\underline{x_l}\} 
\end{equation} 
\begin{equation} \label{eq:10} 
\underline{e}\mathrm{=}\left\{\left.\left[\underline{e_{\mathrm{1}}}\mathrm{,\ }\mathrm{\cdots }\mathrm{,\ }\underline{e_l},\ \cdots \mathrm{,\ }\underline{e_L}\right]\mathrm{\ }\right.\mathrm{|\ }\underline{e_l}\mathrm{\ge }\underline{x_l}\mathrm{,\ }\underline{e_l}\mathrm{<}\overline{x_l}\right\} 
\end{equation} 
\begin{equation} \label{eq:11} 
e_n\mathrm{=}\left\{\left[e_{n_{\mathrm{1}}}\mathrm{,\dots ,\ }e_{n_M}\right]\mathrm{\ |\ }\left[e_{n_{\mathrm{1}}}\mathrm{,\dots ,\ }e_{n_M}\right]\mathrm{\ }\mathrm{\subset }{\mathcal{V}}_N\right\} 
\end{equation} 

\begin{figure}[htbp]
\centering
\includegraphics[width=7cm]{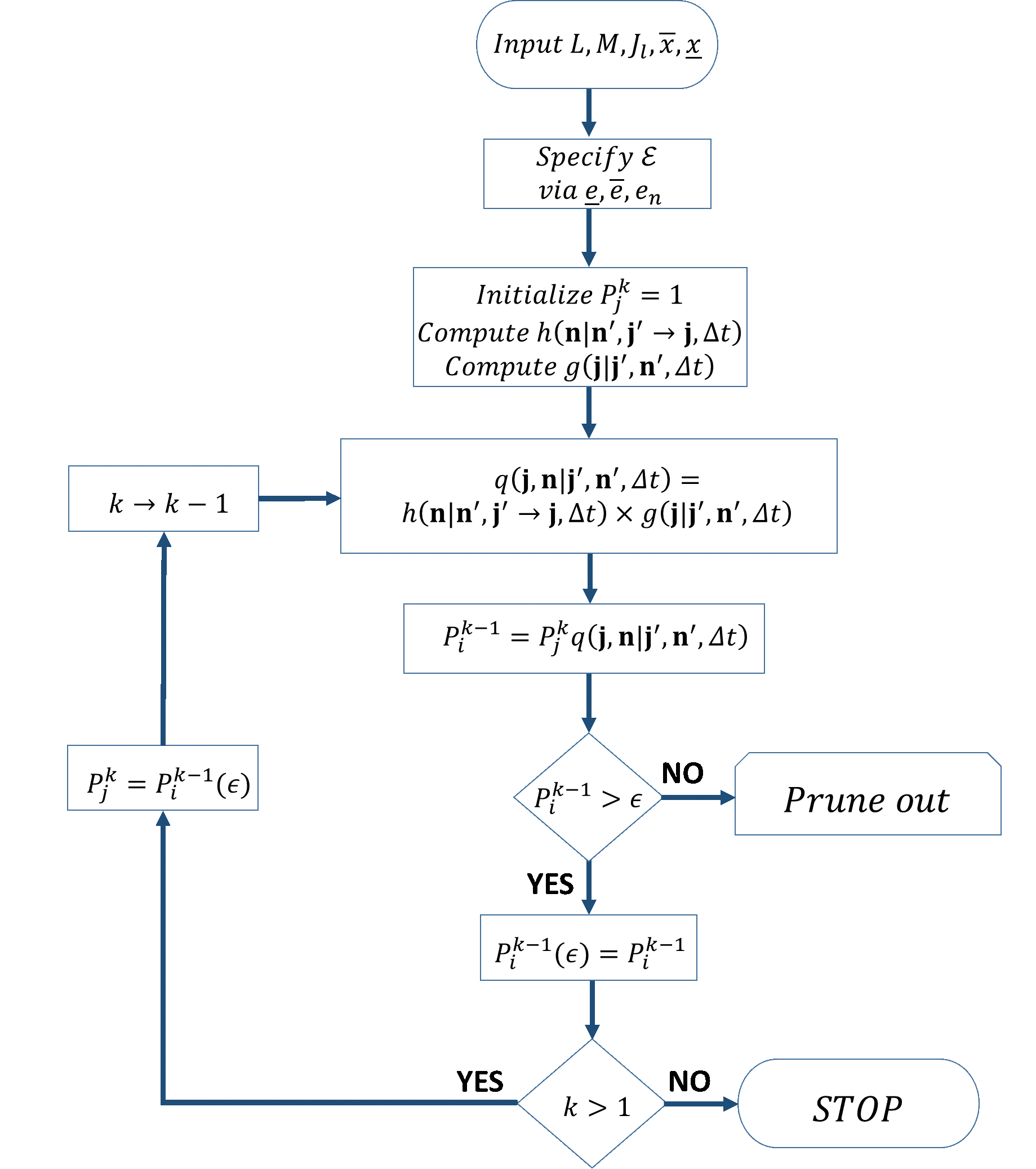}
\caption{Algorithm for Path Probability Calculation (adapted from \cite{yang2016algorithm})}
\label{fig4}
\end{figure}

However, event definition can include a specific system configuration as well, in general. Sequential paths with non-zero transition probabilities that span backwards by a search depth of$\mathrm{\ }k$ time-steps from the event of interest to cells contained in the cell space are then identified. Fig. \ref{fig4} graphically illustrates an algorithm for the BPA. Only the paths with probabilities greater than a user-specified probability truncation value$\mathrm{\ }\epsilon $ are kept in the \textit{Prune Out} process. 

\section{Case Study} \label{CaseStudySection}
\indent 
A high-fidelity simulator for an Autonomous Ground Vehicle (AGV) based on the full 4-wheel model from the work of Ozguner et al. \cite{ozguner2011autonomous} was constructed in Matlab/Simulink environment. The states of the vehicle used in the analysis are the forward velocity, sideward velocity, yaw rate, yaw, \textit{x}-position, and the \textit{y}-position. The control surfaces that affect the vehicle are the engine traction force, the braking force, and the steering angle. Vehicle parameters were taken to be those of the 2009 Lincoln MKS. 

A Hybrid State Control System was used for the decision making and control of the AGV. Two environments are taken into consideration, an urban environment, and a highway environment, with various phases modeled for each of the two environments. The design procedure found in \cite{hejase2016hierarchical} was used for the design and construction of the Hybrid State Control System. Low-level controllers were designed using LQR controllers to control the body rates, and a PI controller to control Euler positions and angles. A FSM that serves as a high-level controller which guides the AGV through the different phases of possible scenarios was constructed. 

\begin{figure}[htbp]
\centering
\includegraphics[width=10cm]{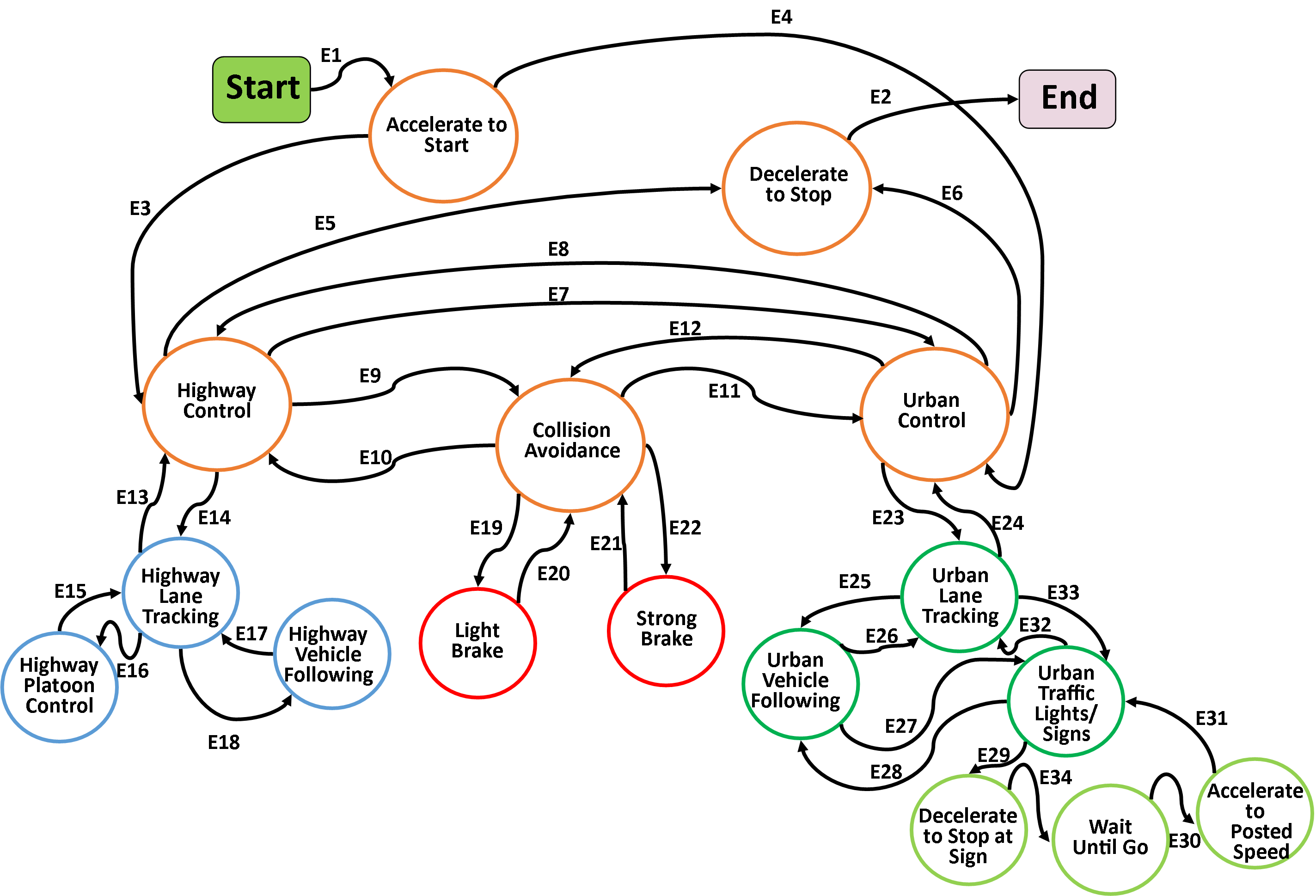}
\caption{Autonomous Ground Vehicle Finite State Machine for High-Level Decision Making}
\label{fig5}
\end{figure}

The hazard of interest (i.e. Top Event) is taken to be a collision with a stationary vehicle in an urban environment. This hazard was selected and modeled based on a list of selected pre-crash scenarios published by the U.S. Department of Transportation \cite{najm2007pre}. For the sake of simplicity in illustrating BPA, it is assumed that in such a scenario the brake-by-wire is the only system component that is prone to random hardware failures. Based on the hazard of interest, the FSM was augmented with emergency and contingency actions for collision avoidance. The resulting FSM can be seen in Fig. \ref{fig5}. The contingency actions that were augmented to the FSM are based on the work of \cite{park2014hybrid}.

Three `brake condition states' were defined in this case study as seen in Table \ref{tab:BrakeMatrix}. The first is Brake Normal, in this state the brake operates normally. The second state is `Minor Brake Fault', in this state the braking system delivers 50\% of what the controller asks of it. The third state is a `Major Brake Fault', in this state the braking system only delivers 25\% of what the controller asks of it. Each of the failed brake states is assigned a probability of $\mathrm{\ }\mathrm{\lambdaup }\mathrm{=(2}\mathrm{\times }{\mathrm{10}}^{\mathrm{-}\mathrm{7}})/h$ which is fairly consistent with some of component failure probabilities in literature [34]. The failures are assumed to be permanent ones.

\begin{table}[htbp]
\caption{Brake States Transition Probabilities}
\label{tab:BrakeMatrix}
\centering
\vspace{0.1cm} 
\hspace{6cm} 
Final Brake State\\
\begin{tabular}{lllll}
\cline{3-5}
\vspace{0.1cm}
& & Normal State & Minor Fault & Major Fault \\ \cline{1-5}
Initial Brake & Normal State (N) & $\approx$ 1 & 2 x ${10^{-7}}$ /h &  2 x ${10^{-7}}$ /h\\
State & Minor Fault & 0 & 1 & 0 \\
& Major Fault  & 0 & 0  & 1 \\
\end{tabular}
\end{table}

The host AGV is initially assumed to be on the road at a positon of (0, 0) in a single lane urban environment with a posted speed limit of 15m/s ($\mathrm{\approx}$35mph). It is also assumed that the stationary target vehicle, which the AGV has to avoid a collision with, is at a position of x=500m at all times. It is assumed that the vehicle is equipped with a sensor that allows it to sense and detect other vehicles that are within a range of 100m ahead. 

The AGV is initially in the urban lane tracking state, and then encounters a target vehicle within a range of 100m. The host vehicle then switches to the urban vehicle following state and aims to follow the target vehicle at a desired time-gap of 1.3s, such that a desired clearance is obtained from$\ c_{des}=t_{gapdes}\times v_{host}$, and is 20m. Once the time-gap from the target vehicle is less than the desired time-gap (i.e.$\ c_{h-t}<c_{des}$), the vehicle switches to collision avoidance and comfortably brakes at -0.3g. If the time-gap from the target is within less than half the desired time-gap (i.e.$\ c_{h-t}<\frac{1}{2}c_{des})$, the vehicle applies a strong brake at -0.8g.

\begin{figure}[htbp]
\centering
\includegraphics[width=10cm]{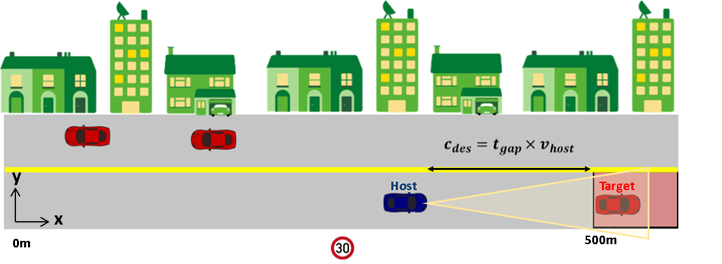}
\caption{Illustration of Autonomous Vehicle Scenario in the Proposed Case Study}
\label{fig6}
\end{figure}

The event, or hazard of interest, is when the AGV reaches an x-position of 500m or greater. This would indicate that a collision took place with the target vehicle. An illustration of the constructed scenario with the hazard of interest can be seen in Fig. \ref{fig6}. 

The aim of this analysis is to identify the risk significant scenarios that lead to a hazard, or a violation of the safety goal. Emergency and contingency actions can then be modified based on the identified scenarios. This process can then be iteratively used to modify contingency actions, until results ensure that scenarios only lead to the violation of a safety goal within acceptably low probabilities.

\section{Simulation} \label{SimulationSection}
\indent

\begin{figure}[htbp]
\centering
\includegraphics[width=16cm]{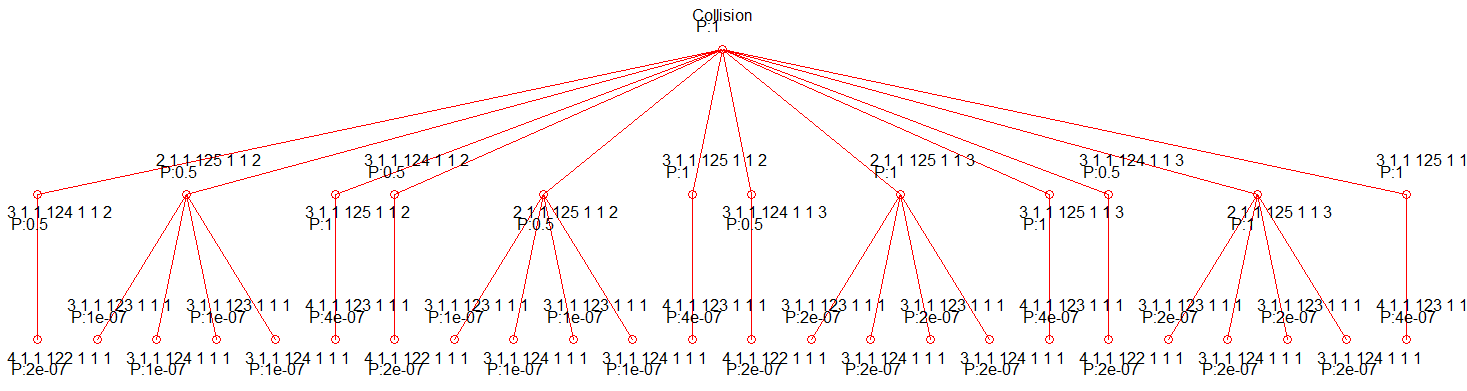}
\caption{BPA results with truncation at 1e-8}
\label{fig7}
\end{figure}

\begin{figure}[htbp]
\centering
\includegraphics[width=8cm]{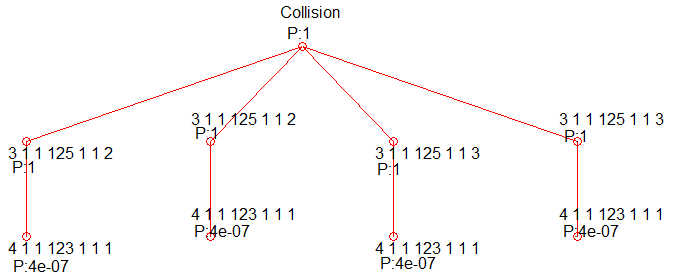}
\caption{BPA results with truncation at 3e-7}
\label{fig8}
\end{figure}

Results based on user inputs from Table III to BPA over a search depth of$\ 2\mathrm{\Delta }t$ with a truncation of scenarios occurring with probabilities $<{10}^{-8}$ can be seen in Fig. \ref{fig7}. Each time step was taken to be 2/3 seconds in length. Note that a search step of two time steps was selected since they amount to 1.3 seconds, this is the time-gap at which the contingency actions begin. Another search tree was also constructed in Fig. \ref{fig8} to illustrate truncation of scenarios occurring with probabilities$\ <{3\times 10}^{-7}$. The truncation probability is used to remove risk insignificant values from the search tree. The probabilities displayed in the search tree are used as probabilistic metrics that rank risk significance of scenarios in comparison to one another. Recall from Section \ref{CaseStudySection} that system at hand has six continuous states (i.e., forward velocity, sideward velocity, yaw rate, yaw, \textit{x}-position, and the \textit{y}-position, and one system hardware configuration (i.e. brake-by-wire). This means that each cell in the discretized space is represented by 7 integers. The first 6 integers represent the segment number of the partitioned continuous variables based on the system discretization, and the 7${}^{th}$ integer represents the brake condition. Each node in the tree of Figs. 7-8 contains 7 integers and an associated probability. 

\begin{table}[htbp]
\caption{User Input to BPA}
\label{tab:userInput}
\centering
\vspace{0.05cm}
\begin{tabular}{lll}
\cline{1-3}
\end{tabular}
\vspace{0.05cm}
\begin{tabular}{lll}
\cline{1-3}
\vspace{0.1cm}
Variable Name & Notation & Value\\
numProcessVariables & L & 6 \\
processVariablesNames &  & ["Fwd Vel.", "Side Vel.", “Yaw Rate”, “x-Pos”, “y-Pos”, ”Yaw”] \\
numSystemComponents & M & 1 \\
systemComponentNames & & ["Brake State"] \\
systemComponentStates & ${N_i}$ & [3] \\
systemComponentStateNames & & [“Normal”, “Minor Brake Fault”, “Major Brake Fault”] \\
variableUpperBounds & \(\overline{x}\) & [20,5,0.5,600,6,pi/3] \\
variableLowerBounds & \(\underline{x}\) & [0,-5,-0.5,0,-6, 0-pi/3] \\
numberOfCells & ${J_i}$ & [5,1,1,150,1,1,3] \\ 
sysConfTransProb & ${H_{n_m}}$ & 
\(\begin{bmatrix}
    \approx 1 & 2e-7 & 2e-7 \\
    0 & 1 & 0 \\
    0 & 0 &  1
\end{bmatrix}\) \\
eventUpperBounds & \(\overline{e}\) & [20,0.5,0.5,600,6,pi/3,3] \\
eventLowerBounds & \(\underline{e}\) & [0,-0.5,-0.5,500,-6, 0-pi/3,1] \\
\end{tabular}

\end{table}

Taking the first sequence from the left in Fig. \ref{fig7} as an example sequence to interpret results from branch the search tree,
\begin{equation} \label{eq:12} 
{\left[\mathrm{4\ 1\ 1\ 122\ 1\ 1\ }\boldsymbol{\mathrm{1}}\right]}_{P\mathrm{=2\times }{\mathrm{10}}^{\mathrm{-}\mathrm{7}}}\mathrm{\to } {\left[\mathrm{3\ 1\ 1\ 124\ 1\ 1\ }\boldsymbol{\mathrm{2}}\right]}_{P\mathrm{=0.5}}\mathrm{\to }Collision 
\end{equation} 
we can make the following observations:
\begin{enumerate}
\item  $\left[\mathrm{4\ 1\ 1\ 122\ 1\ 1\ }\boldsymbol{\mathrm{1}}\right]$-- The AGV initially has a forward velocity of 12 to 16 m/s, a sideward velocity of -0.5 to 0.5m/s, a yaw rate of -0.5 to 0.5 rad/s, an x-position of 488-492, an y-Position of -6 to 6m, and a Yaw angle of --$\piup$/3 to $\piup$/3. The brake state was Normal. 

\item  $\left[\mathrm{3\ 1\ 1\ 124\ 1\ 1\ }\boldsymbol{\mathrm{2}}\right]$-- One time step later, the AGV had a forward velocity of 20 to 25 m/s, a sideward velocity of -0.5 to 0.5m/s, a yaw rate of -0.5 to 0.5 rad/s, an x-position of 496 -- 500m, an y-Position of -6 to 6m, and a Yaw angle of --$\piup$/3 to $\piup$/3. The brake experienced a Minor Brake Fault.

\item      \textbf{Collision} -- One time step later the AGV collides with the stationary vehicle located at an x-position that is greater than 500m, leading to a violation of the safety goal. 
\end{enumerate}

Upon investigation of the sequences in the search tree of Fig. \ref{fig7}, it is also observed that the safety goal is violated once a brake failure occurs. In an effort to address this observation, the contingency actions were modified. The time-gap was changed to 2s, such that a Light Brake contingency action is employed once the host vehicle is within 30m clearance from the target vehicle, rather than 20m. The Strong Brake contingency action is employed once the host vehicle is within 15m of the target vehicle, rather than 10m. Based on these modifications in the contingency actions, with all other parameters from Table \ref{tab:userInput} kept the same. 
The BPA was run again over a search depth of $3\mathrm{\Delta }t$ and truncation of branches with probabilities$\ <{10}^{-8}$. A search depth of $3\mathrm{\Delta }t$ was selected since this amounts to 2 seconds, the time-gap at which contingency actions begin. No paths of risk significance leading to the Top Event were identified, which means that the proposed contingency actions managed to bring the system to a safe state within an acceptable risk level, even under the occurrence of the assumed hardware failures. This example also illustrates how BPA can be used towards a safer design.

\section{Challenges Faced in Automotive Scenarios, and Potential Solutions} \label{ChallengesSection}
\indent

The ultimate goal of the described methodology is the design of a generic quantitative risk assessment scheme that is capable of providing information on risk-significant sequences of events that violate safety goals. 
Safety assurance of control systems being developed for automotive scenarios has multiple challenges. Two main challenges are identified by the authors: 
1) Large-scale scenarios that involve high levels of autonomy and many hardware components do not typically have a single domain expert that is able to accurately set-up BPA parameters for the overall scenario. 
2) For autonomous systems with a large state-space, such as platoons of vehicles, combinatorial and computational issues are prone to appear. 

Future work of BPA is directed towards solving the identified challenges. A possible solution to Challenge 1 is running phase-specific implementations of BPA, and integrating results of analysis obtained from the multiple phases. The authors are already in the process of developing a generalized scheme for such a solution, with a preliminary approach described in \cite{hejase2018dynamic}.
The nature of BPA equips it with tools that can naturally help alleviate problems faced due to Challenge 2. Through selection of larger cell sizes (a more coarse partitioning scheme), and sampling more cells from each cell, the size of the system cell-to-cell map can be reduced. Noting that the reduction in size is compensated by sampling more points from each cell. This in turn reduces the computations needed to identify risk significant path sequences. Additionally, careful and intelligent selection of the truncation value parameter can lead to reduced wastage of computational resources on risk insignificant event sequences.

\section{Conclusion} \label{ConclusionSection}
\indent

The need for generic and well defined procedures and methods for the assurance of autonomous ground vehicle functions with respect to safety goals in the early design stage is of vital importance. In this paper, the BPA approach based on a deductive implementation of Markov/CCMT has been proposed for the identification of scenarios that lead to safety goal violations. The scenarios were ranked by risk significance via probabilistic quantification of the scenarios that violate the safety goals. A case study of a hybrid state autonomous vehicle prone to random hardware failures in the braking system was taken into consideration. Simulation results displayed the risk significant scenarios leading to collisions with a static vehicle under possible brake failure in a search tree format. The simulated scenarios indicated that even though the contingency actions work as required under nominal brake conditions, they did not adequately avoid the risk of collision under sub-nominal brake conditions.  
It was also shown that, based on the modification of the contingency actions, no paths of risk significance leading to the Top Event were identified. 
Future work will involve the definition of a broader and more realistic set of contingency actions and hardware failures for various phases of the scenario. Future work will also investigate the applicability of the BPA to current standards such as ISO 26262 \cite{iso201126262}. As a final note, it should be indicated that while BPA, in principle, may lead to combinatorial increase in the number of scenarios to be investigated, an upper limit can be imposed on this number through the specification of probability bounds in defining what is regarded as risk significant. This probability bound can be relaxed in new BPA runs once the initially identified scenarios of high risk significance are mitigated.

\section*{Acknowledgement}
\indent 
The work is partially funded by the National Science Foundation (NSF) Cyber-Physical Systems (CPS) project under contract 60046665. An application BPA to Unmanned Aircraft Systems (UAS) was developed with ASCA Inc. as part of a project funded by the NASA Ames Research Center (ARC). Discussions in that project with Drs. Sergio Guarro, and Michael Yau from ASCA Inc., and Dr. Matt Knudson from NASA ARC are gratefully acknowledged. 

\nocite{*}
\bibliographystyle{eptcs}
\bibliography{SCAV2018Ref}

\end{document}